# Solid phase of Krypton on the exterior of individual single-walled carbon nanotubes.

Oludotun B. Ode and Silvina M. Gatica

*Department of Physics and Astronomy, Howard University, 2355 6th St. NW, Washington DC 20009*

We have computed the adsorption of Krypton in a closed single-walled carbon nanotube using the method of Grand Canonical Monte Carlo. Our results indicate evidence of an incommensurate solid formed at high pressure and low temperature ($T$ < 85 K), before the formation of a second layer. The solid melts above that temperature. Our simulations are in good agreement with novel experimental results for adsorption in individual carbon nanotubes.

1. INTRODUCTION

The investigation of adsorption phenomena has been a very exciting and successful scientific activity in the last half-century. In recent years a considerable amount of work has been devoted to investigate theoretically and experimentally the physical adsorption in bundles of carbon nanotubes (NTs).[1,2,3,4,5]

Recently obtained experimental results focus the attention on individual carbon nanotubes. Data of adsorption of Kr on a *single* nanotube, from the Cobden-Vilches collaboration at the U. Washington,[6] show two quite remarkable, nearly vertical transitions. Those steps are observed and interpreted as cylindrical surface analogues of 2D monolayer transitions: vapor to commensurate solid (CS) while the second jump is to an incommensurate solid (IS) phase coating the nanotube's surface. Explaining this behavior and predicting that of other gases (only Ar and Kr have been studied experimentally thus far) requires the understanding of the adsorption in individual nanotubes. In our simulations we find that steps in the isotherms are present, due to the transition from vapor to an incommensurate solid, in agreement with Vilches' observation. In this work we did not search for the commensurate solid phase, since we used a continuous approximation for the nanotube that neglects the corrugation of the wall. The CS phase is currently being investigated in our group and will be reported in the near future.

Kr is known to form CS and IS phases on top of flat graphite [7]. Similar phases are expected to form on the surface of a NT, provided that the radius ($R$) is large enough. For smaller radii, the increasingly important effect of the curvature would eventually alter qualitatively the phase diagram preventing some phases and/or allowing new ones. Lueking and Cole investigated this interesting problem. [8] Their findings for Kr show that the CS phase remains stable at 0 K on nanotubes of $R$ > 0.7 nm, while for narrower tubes the so called "striped phase" is the most stable one. Similar results are found for other gases.

In this article we investigate the possible IS phase. We have performed Grand Canonical Monte Carlo simulations using a continuous model for the potential of the nanotube.

This paper is organized as follows. In the next section we present our method of computation, in Section 3 we discuss our results for Kr in the exterior of a NT. The last section is dedicated to a summary and conclusions.

## 2. METHOD OF COMPUTATION

As in most adsorption studies on graphite or carbon nanotubes, the adsorption potential used here is a pair-wise sum of two-body interactions $U(\mathbf{x})$ between a molecule and the nanotube's carbon atoms. The pair potential is assumed to be isotropic and of Lennard-Jones (LJ) form:

$$U(r) = 4\varepsilon\left((\sigma/r)^{12} - (\sigma/r)^{6}\right), \quad [1]$$

where $r$ is the interatomic distance and the LJ parameters $\varepsilon$ and $\sigma$ for the Kr – C interaction are obtained with semiempirical combining rules

$$\sigma_{KrC} = (\sigma_C + \sigma_{Kr})/2 \quad [2]$$

$$\varepsilon_{KrC} = \sqrt{\varepsilon_C \varepsilon_{Kr}} \quad [3]$$

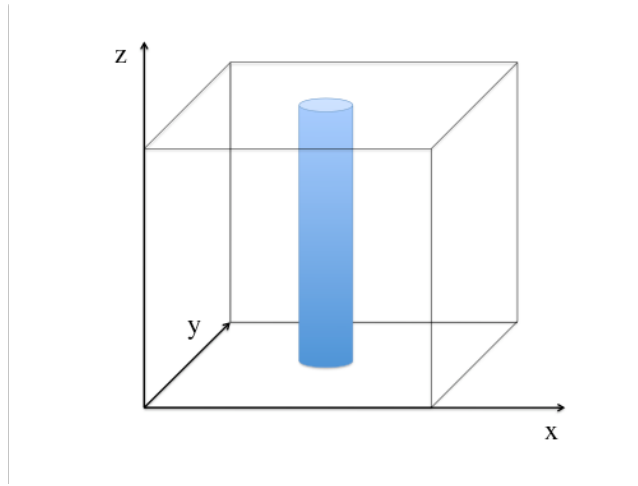

FIG 1. Schematic view of the simulation cell, not in scale. The sides are 6.8 nm in the $z$ direction and 7.2 nm in $x$ and $y$.

The parameter values used in this study are $\varepsilon_C$ = 28 K, $\varepsilon_{Kr}$ = 171 K, $\sigma_C$ = 0.34 nm and $\sigma_{Kr}$ = 0.36 nm.[9] Another simplifying assumption employed here is the replacement of discrete carbon atoms by a continuous cylindrical sheet of matter; this approximation, although it prevents us from finding any commensurate phase, is a suitable model for the study of incommensurate phases. [9,10]

We have performed GCMC simulations of Kr interacting with the outer wall of a single nanotube. In the GCMC simulations, the chemical potential, temperature, and volume are held constant while the number of particles varies. This technique is standard and we refer the reader to refs. [9, 10] for details. For a single isotherm point typically, 3 million moves were performed to equilibrate the system and 1 million moves were used for data gathering. The Kr–Kr pair interactions were taken as a 12-6 Lennard-Jones potential as well. The nanotube was positioned at the center of the simulation cell with the axis in the z direction. The size of the simulation cell was carefully chosen verifying that the results were unaffected by the position of the walls. We simulated an infinitely long tube by setting periodic boundary conditions in the *z* direction. The box's walls perpendicular to the *xy* plane were set reflective. (See Fig. 1 for a schematic description).

3. RESULTS

Our results are shown in figures 2 to 6. Although the study reported in this paper corresponds to the particular case of a 0.7 nm-radius nanotube, the behavior is not exclusive of this particular choice as other sizes were studied with qualitatively similar outcome.

Figure 2 displays the adsorption isotherms at different temperatures going from 73.7 K to 120 K. Here we present the coverage in units of areal density, i.e. number of krypton atoms per unit area of the cylindrical shell. The radius of the shell is 1.06 nm. For the lowest temperature shown in the figure, the data shows an abrupt step in the uptake going from almost zero to 5.7 $nm^{-2}$ and then increases continuously to 6.7 $nm^{-2}$. The top value is smaller than the areal density of a 2D triangular lattice of spacing equal to 0.4 nm (2D LJ Kr), that is 7.2 $nm^2$.

This transition from vapor to a monolayer is observed up to temperatures as high as 90 K. At higher temperatures, the isotherms are smoothed out by thermal effects, as it is usually the case in physical adsorption. The monolayer phase is a cylindrical surface surrounding the exterior of the nanotube, coaxial with the tube, as seen in Figure 3. The radius of the Kr shell is 1.06 nm, implying that the distance to the carbon surface is 0.36 nm. Further increasing the pressure would eventually produce a second layer or multi layers, that appear as a second sharp step up. The monolayer phase extends over a wide range of pressure, going from 0.03 Torr to 1.64 Torr (for *T* = 73.7 K). At *P* = 0.14 Torr, a small step is observed, where the structure of the layer changes from liquid-like to solid-like. The solid-like phase has the aspect of a triangular lattice, if the cylinder is unrolled into a plane, as the one seen in Fig. 4 (for 30 K). Thus, three transitions are observed at 73.7 K: vapor to liquid (V-L), L to incommensurate solid (L-IS) and IS to bilayer (S-BL). Similar transitions are found at 77.4 K.

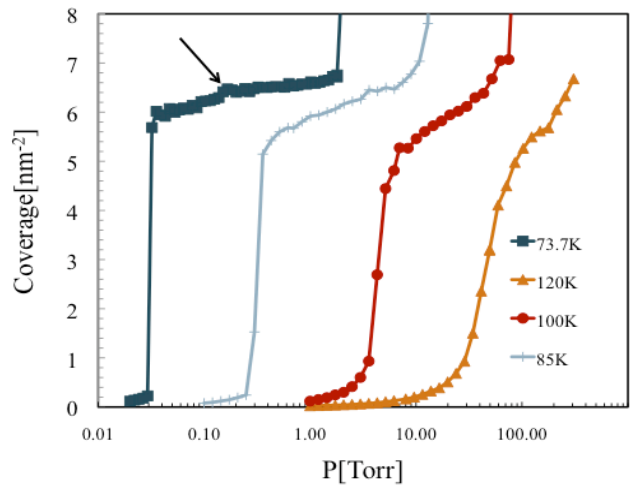

Fig. 2. Adsorption isotherms for Kr in a closed 0.7 nm radius nanotube for temperatures shown in the legend. The arrow indicates a possible L-IS transition. The lines are a guide to the eye.

At low temperatures (30-40K), the liquid phase is absent and only two transitions occur (V-IS, IS-BL). At higher temperatures (85K-90K), there is a sharp step that corresponds to V-L transition and then the coverage grows continuously to a multilayer film, with no evidence of a IS phase. Finally, for $T>100K$ the coverage grows continuously.

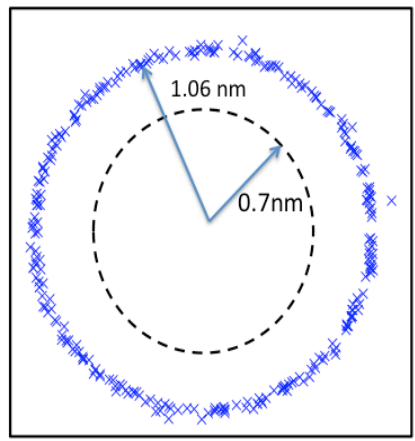

Fig. 3. Projection of the position of Kr atoms on the $xy$ plane, perpendicular to the axis of the nanotube, at T = 73.4 K and P = 0.1 Torr. The dashed circle represents the surface of the nanotube. The atoms spread over a range of 7.2 nm in the $z$ direction.

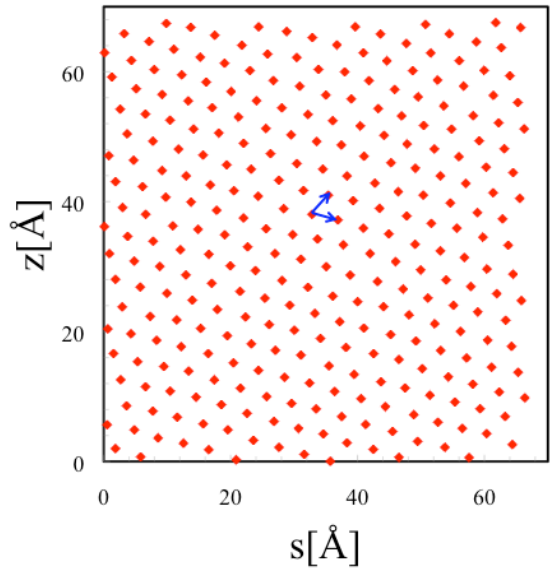

Fig. 4  Coordinates of the Kr atoms for one of the equilibrium configurations at $T$ = 30 K and $P$ = 8.6 ×10$^{-13}$ Torr. The coordinates $z$ and $s$ are in the direction of the axis of the nanotube and the circular arc, respectively. The arrows indicate the primitive lattice vectors of size 0.4 nm (up) and 0.45 nm (down).

In the region of the L-IS transition it appears to be a small step in the isotherms, that may indicate a phase transition. However, in order to conclude that, a coexistence of phases should be present.  Efforts to include an evaluation of the structure in the monte carlo are in progress.  Here we report the structural study of random equilibrium configurations at different pressures and temperatures. From the visualization of the position of the atoms we can estimate the degree of order in the structure. An example for $T$ = 30 K is displayed in Fig. 4. Here, the position of the Kr atoms are plotted on the unrolled plane, for clarity.  In this way we are able to observe a lattice analogous to the IS phase that Kr forms on top of flat graphite[7]. The lattice is tilted with respect of the horizontal, as a consequence of the geometry constraining the lattice to "fit" around a cylinder. Also, the lattice is slightly stretched in the direction perpendicular to the axis, due to the curvature of the surface.

To further investigate the structure of this phase we calculated the two-dimensional "radial distribution function" (RDF), defined as

$g(r) = N_r/(2\pi r\, dr\, \theta)$        [4]

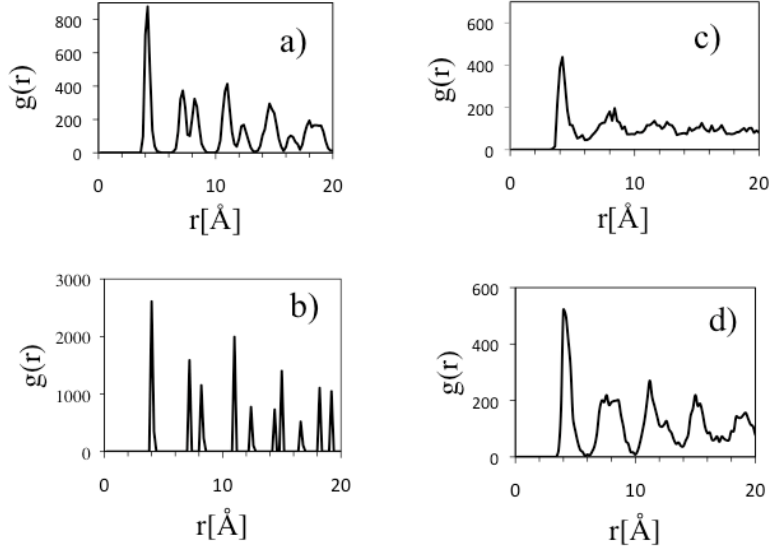

FIG 5. Radial distribution function for sample configurations: a) at $T$ = 30 K and $P$ = $1.36 \times 10^{-12}$ Torr, b) for a regular triangular lattice, c) and d) at $T$ = 73.7K, $P$ = 0.035 Torr and 0.069 Torr respectively.

where $N_r$ is the number of pairs of atoms at a distance $r - r+dr$ in the unrolled plane and $\theta$ is the number of atoms per unit area. The results for a configuration at 30 K are displayed in Figure 5a). There is a well defined main peak located at $r$ = 0.42 nm, and 0.12 nm wide. For comparison we also show the RDF of a regular triangular lattice of lattice constant $a$ = 0.42 nm. The structure of the lattice is not lost in the configuration corresponding to Fig. 5a), although the peaks are wider. Figs. 5c) and 5d) correspond to configurations in the monolayer phase at $T$ = 73.7 K and low and high pressures respectively. In the case 5d), the structure resembles the solid-like structure of 5a) and b), opposite to case 5c) where the RDF has a liquid-like aspect. There is clearly a qualitative difference between 5d) and c). Notice that the position of the main peak in all four figures is at $r$ = 0.42 nm. This value is 5% larger than the minimum of the Kr-Kr potential, 0.4 nm. This shift is due to the construction done to visualize the structure, where the cylindrical shell is mapped into the unrolled plane. This also causes the peak to be wider since the distortion occurs only in the direction perpendicular to the axis. The two next secondary peaks correspond to the second nearest neighbor (0.72 nm) and twice the lattice constant (0.83 nm). Several other peaks follow as seen in Fig. 5.

For increasing temperatures, the peaks become wider and the structure function loses the solid-like features indicating the melting of the Kr layer.

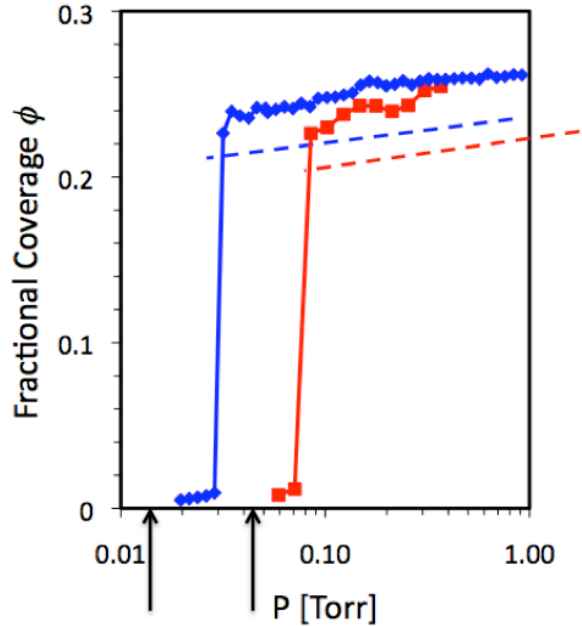

FIG. 6. GCMC isotherms at $T = 73.7$ K (blue diamonds) and $T = 77.4$ K (red squares) compared to experimental results from Wang et. al. The dashed lines represent the experimental IS phase at 73.7 (blue, left upper) and 77.4 (red, right lower), extracted from Fig. 3 in Ref [6]. The arrows indicate the experimental uptake pressure at 73.7 K (left) and 77.4 K (right), from the same figure in Ref. [6]

The comparison of our data with the experimental results of Wang et. al. [6] are presented in Fig. 6. See also Fig.3 in Ref. [6] for a complete picture of the experimental results. Here we plot the *fractional coverage* $\phi$ defined as the number of adsorbed Kr atoms per carbon atom. Our simulation data at 73.7 K and 77.4 K have one abrupt step only, and differ from the experimental data in three main aspects. First, the experimentally found CS is not evident in the simulation. This was expected since our simulation uses a continuous approximation for the nanotube. Secondly, in the simulation the first step occurs at a higher pressure than in the experimental data. This may be explained by a mismatch between the size of the nanotube used in the simulation and the one in the experiment as follows. The radius of the nanotube in the experiment is not known,[11] while the radius used in the simulation is 0.7 nm. From the comparison we may deduce that the nanotube in the experiment would be wider, since the attraction increases, and hence the uptake pressure decreases, with decreasing curvature. In Fig. 7 we show the effect of the curvature on the uptake pressure $P_u$ computed with the GCMC method, for $T = $ 73.7K. From this curve, and taking the experimental value $P_u = 0.013$ Torr from Ref.

[6], we estimate the radius of the nanotube used in the experiment to be 1.3 nm. It is also possible that the model we are using for the potential interaction underestimates the attraction of the nanotube. We are currently investigating this possibility. The third main difference is that the fraction of coverage corresponding to the IS phase in the simulation is approximately 10% larger than in the experiment. This discrepancy may be also attributed to the diferent radii of the nanotubes used in the simulation and the experiment, as we explain in the following. The fractional coverage $\phi$ is defined as the number of krypton atoms adsorbed divided by the total number of carbon atoms in the nanotube, and can be expresed as,

$$\phi = \frac{R_{Kr}\theta_{Kr}}{R_{NT}\theta_{NT}} \qquad [5]$$

in therms of the areal densities and radii of the carbon nanotube and Kr shell. If the radius of the nantoube increases, so it does the radius of the Kr shell that acording to our simulation is at a distance 0.36 nm from the carbon surface, i.e. $R_{Kr} = R_{NT} +$ 0.36nm. Assuming that this distance remains unchanged for a larger radius, we can calculate from eq. 5 the ratio of the coverages on a 0.7 nm nanotube to that on a 1.3-nm nanotube, $\phi_{0.7nm}/\phi_{1.3nm} = 1.19$. Although this value overestimates the observed effect, it gives the right trend.

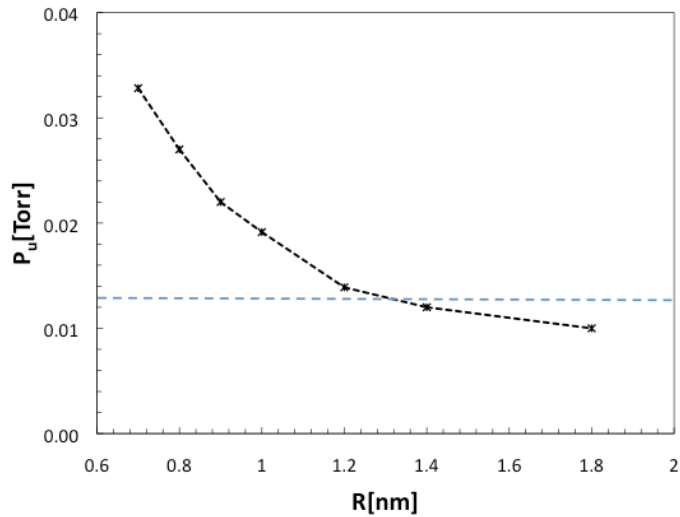

FIG. 7. Uptake pressure $P_u$ as a function of the radius of the nanotube, from GCMC simulations at $T$ = 73.7 K. The dashed line represents the experimental value, from Ref. [3].

4. SUMMARY

In this work we have studied the adsorption of Krypton on the outer surface of an individual single-walled carbon nanotube. Our work was inspired by recent experimental results reported in Ref. 6.  We have found that Kr forms an incommensurate solid phase analogous to the one found on flat graphite. The comparison with the experiment is satisfactory, as the discrepancies can be understood in terms of a mismatch in the size of the tubes used in the experiment and the simulation.  However we emphasize that the model used here for the potential is not suitable to investigate the commensurate solid phase. [12]

Although most of the results reported here correspond to simulations done for a 0.7-nm "smooth" nanotube, we have explored other sizes and the atomistic model without finding significant differences.  In any case, the phase transition we investigate here would not be qualitatively affected by the size of the nanotube, since it is an "incommensurate" phase.

Acknowledgments

We are grateful to Milton Cole, Louis Bruch, Oscar Vilches, Zenghui Wang and David Cobden for helpful comments and discussions.

This research has been supported by Howard University New Faculty Start-Up award. This research used resources of the National Energy Research Scientific Computing Center, which is supported by the Office of Science of the U.S. Department of Energy under Contract No. DE-AC02-05CH11231. O. O. acknowledges the support of the Howard University/Johns Hopkins University/Prince George's Community College Partnership for Research and Education in Materials, NSF Award 0611595.

REFERENCES

[1] M. Mercedes Calbi, Milton W. Cole, Silvina M. Gatica, Mary J. Bojan and J. Karl Johnson, "Adsorbed gases in bundles of carbon nanotubes: theory and Simulation", Chapter 9 of Adsorption by Carbons , edited by E. J. Bottani and Juan M. D. Tascón, Elsevier Science Publishing, pp. 187-210 (2008)

[2]  A. D. Migone, "Adsorption on carbon nanotubes: experimental, "Adsorption on carbon nanotubes: experimental results", in "Adsorption by Carbons", ed. by E. J. Bottani and J. M. D. Tascón, Elsevier (2008)

[3] S. M. Gatica, M. M. Calbi, R. D. Diehl and M. W. Cole, J. Low Temp Phys 152, **89** (2008)

[4]  Colloquium:  Condensed phases of gases inside nanotube bundles, M. M. Calbi, M. W. Cole, S. M. Gatica, M. J. Bojan and G. Stan, Rev. on Modern Phys. **73**,  857 (2001)

[5]  S. Rols, M. R. Johnson, P. Zeppenfeld, M. Bienfait, O.E. Vilches and J. Schneble, Phys. Rev. B **71**, 155411 (2005)

[6] Zenghui Wang, Jiang Wei, Peter Morse, J. Gregory Dash, Oscar E. Vilches, David H. Cobden, Science **327**, 5965, pp. 552 – 555 (2010)

[7] L. W. Bruch, M. W. Cole and E. Zaremba, Physical adsorption: forces and phenomena (Dover, Mineola NY, 2007)

[8] Angela D. Lueking and Milton W. Cole, Phys. Rev. B **75**, 195425 (2007)

[9] G. Stan, M. J. Bojan, S. Curtarolo, S. M. Gatica and M. W. Cole, Phys. Rev. B **62**, 2173 (2000)

[10] S. M. Gatica, M. J. Bojan, G. Stan and M. W. Cole, J. Chem. Phys. **114**, 3765 (2001)

[11] O. Vilches, private communication

[12] S.M. Gatica and M.W. Cole, submitted to the J. of Low Temp. Phys.